\documentstyle[11pt]{article}
\include{epsf}

\begin{document}
\thispagestyle{empty}
\begin{flushright}
hep-th/9801098
\end{flushright}
\begin{center}
\vspace*{7mm}
{\Large Loops, Surfaces and Grassmann Representation \vskip2mm
in Two- and Three-Dimensional Ising Models}
\vskip14mm
\centerline{ {\bf
C.R. Gattringer, S. Jaimungal and G.W. Semenoff}}
\vskip 2mm
\centerline{Department of Physics and Astronomy,}
\centerline{University of British Columbia, Vancouver B.C., Canada}
\vskip28mm
\begin{abstract}
Starting from the known representation of the partition function 
of the 2- and 3-$D$ Ising models as an integral over Grassmann
variables, we perform a hopping expansion of the corresponding Pfaffian.
We show that this expansion is an exact, algebraic representation of 
the loop- and surface expansions (with intrinsic geometry) of the 2-
and 3-$D$ Ising models. Such an algebraic calculus is much simpler to deal
with than working with the geometrical objects. 
For the 2-$D$ case we show that the algebra of hopping generators 
allows a simple algebraic treatment of the geometry factors and 
counting problems, and as a result we obtain the corrected loop expansion of 
the free energy. We compute the radius of convergence of this expansion 
and show that it is determined by the critical temperature.
In 3-$D$ the hopping expansion leads to the surface representation 
of the Ising model in terms of surfaces with intrinsic geometry. 
Based on a representation of the 3-$D$ model as a product of 2-$D$ models
coupled to an auxiliary field,
we give a simple derivation of the geometry factor which prevents
overcounting of surfaces and provide a classification of possible
sets of surfaces to be summed over. For 2- and 3-$D$ we derive a compact 
formula for 2$n$-point functions in loop (surface) representation.
\end{abstract}
\end{center}
\vskip2mm
\noindent
\newpage
\setcounter{page}{1}
\section{Introduction}
\subsection{Motivation}
The Ising model is widely used for
illustrating concepts in statistical mechanics and field theory. 
In 2-dimensions
it is exactly solvable and provides the classic example of a theory which 
exhibits non-mean-field critical exponents. There are three
representations of the model: That as a magnetic spin system, as a theory 
of random paths \cite{kac}-\cite{books} 
and as a fermionic lattice field theory with Gaussian
action (Grassmann representation) \cite{fradkin,samuel1,itzykson1}. 
The connections between these are interesting, 
since they illustrate a deep relationship 
between dynamics and geometry.  The random paths can be thought 
of as Euclidean world lines of the fermions and in turn as domain 
boundaries in the spin system. 

Even more fascinating are the corresponding relationships in the 3-$D$ Ising model.
The 3-$D$ model is not exactly solvable.  However, it does share some
of the geometrical features of the 2-$D$ model. Its partition function
can be represented as a spin model, as a model of decorated  random
surfaces \cite{sedra}-\cite{distler} and as a fermionic lattice model which 
is no longer Gaussian \cite{fradkin,samuel3,itzykson2}.
There is an intriguing suggestion \cite{polyakov,polyadots}
that the continuum limit of the 3-$D$ Ising model 
could be some sort of non-critical
string theory. The precise form of such a string theory is yet unknown.

Even the representation in terms of lattice surfaces (or loops)
remains to be completely understood. In particular the calculus of surfaces 
(loops) is rather cumbersome to work with. Handling the symmetry and 
geometry factors in the loop and surface representations is 
a non-trivial problem.  It would be desirable to have an
exact algebraic representation of the geometrical objects. In this article
we shall show that such an algebraic representation can be obtained from the 
hopping expansion (see e.g. \cite{MoMu94}) 
of the Grassmann representation of the model.
We obtain several new results which serve to demonstrate the power of this calculus. 

The language of the hopping expansion allows for 
a simple algebraic formulation of the combinatorics of surfaces and loops. 
In particular all geometric
factors are obtained as traces of ordered products of the hopping generators
(= generators of shifts on the lattice). Also, counting problems which
correspond to some symmetries of the geometric objects, such as 
iteration of loops (i.e. the loop runs through its links several times)
can be tackled in a simple manner. As an application of the latter we shall 
give the corrected version of the loop expansion of the free energy 
in the 2-$D$ case. It differs from previous results by extra factors 
for iterated loops. Also the computation of the radius of convergence of 
loop- or surface expansion is a rather intractable problem if one has 
to work with the geometrical objects. For the 2-$D$ case we show that 
our calculus reduces this problem to the computation of the norm of the 
hopping matrix. 

For the 3-$D$ case we derive a representation of the 3-$D$ Ising model
as a product of 2-$D$ Ising models coupled to an auxiliary field. 
Based on this representation we
give an elegant derivation of the geometric factor 
which eliminates overcounting of surfaces. Our approach also allows for a
simple classification of surfaces to be summed over in the surface 
representation. We show that this classification can be reduced to 
a 2-dimensional problem: The surfaces can be characterized by the loops
that emerge as intersections of the surfaces with the coordinate planes
of the lattice. 
There is some freedom in the choice of admissible loops in these planes.  Different
choices lead to different classes of surfaces. For both 2- and 3-$D$
we derive an elegant formula for 2$n$-point functions in terms of loops and
surfaces, respectively.

The paper is organized as follows: In Section 1.2 we review the representations
of the Ising model in terms of loops (2-$D$) and surfaces (3-$D$) and 
introduce our conventions. Section 2 is dedicated to the 2-$D$ case. 
In 2.1 we set up the Grassmann representation and perform the hopping
expansion. This is followed by Sub-section 2.2 where we show how to obtain the 
representation in terms of loops from the hopping expansion and show that the
latter is an exact, algebraic representation of the loop calculus. The radius
of convergence of the loop expansion of the free energy is shown to be 
determined by the critical temperature in 2.3.

Section 3 gives our results for the 3-$D$ case. In 3.1 we first decompose the
quartic term in the Grassmann representation by introducing an auxiliary field 
and set up the hopping expansion. In 3.2 we extract the surface picture from the
hopping expansion, derive the geometric factors and discuss the above mentioned
classification of surfaces. In Sub-section 3.3 we derive the formula for the 
2$n$-point functions in terms of surfaces (loops). The article closes with
a discussion in Section 4.

\subsection{Partition function, loops and surfaces}
Since we will make extensive use of the representation of the 
partition function 
in terms of loops and surfaces we will review these representations here
and introduce our conventions.

The Ising model in terms of spin variables has the partition function 
\begin{equation}
Z \; = \; \sum_{\{ s(x) = \pm 1 \}} 
\exp \left( \beta \sum_{\langle x, y \rangle} s(x) s(y) \right) \; ,
\label{zorig}
\end{equation}
where $x$ runs over all sites of the $D$-dimensional lattice $\Lambda 
=$Z\hspace{-1.3mm}Z$^D$ ($D =2,3)$ and $\langle x, y \rangle$ denotes nearest 
neighbors.

The 2-$D$ partition function has a representation in terms of 
closed loops \cite{kac}-\cite{books},\cite{distler}
on the dual lattice $\Lambda^*$ (for the 
square lattice $\Lambda \simeq \Lambda^*$). 
This representation is obtained by 
drawing closed loops $\gamma$ on the dual lattice 
around patches of negative spins on the original lattice.
Each link in $\gamma$ crosses a link of the original lattice which has
anti-aligned spins at its endpoints. By drawing loops around all such patches
every anti-aligned link is taken into account. Each of 
these links has a Boltzmann weight exp$(-\beta)$, and since the total 
length $|\gamma|$ of all
loops is equal the number of anti-aligned neighbors, this gives rise 
to the weight exp$(-\beta |\gamma|)$. 
If $V$ denotes the number of all sites, there 
remain $2V - |\gamma|$ links with aligned spins at the 
endpoints\footnote{In order to make all intermediate formulas well defined
we formally work on a finite lattice, ignoring boundary terms, since in 
the end we perform the thermodynamic limit.}. They
have Boltzmann weight exp$(+\beta)$ and one finds 
\begin{equation}
Z \; = \; 2 \; t^{-2V} \sum_{\gamma \in {\cal L}_{ext} } 
(t^2)^{|\gamma|} \; .
\label{z2ext}
\end{equation} 
Here we introduced $t =$ exp$(-\beta)$.
The factor 2 emerges, since every loop configuration corresponds to 2
spin configurations related by the Z\hspace{-1.3mm}Z$_2$ symmetry of the
model. ${\cal L}_{ext}$ is defined to be the set of 
closed loops (not necessarily connected)
which can be obtained by drawing lines on the dual lattice around patches of
negative spins, so that a curve $\gamma$ in ${\cal L}_{ext}$ is  
given by a collection of links that have no boundary. 
This defines loops by their so-called {\it extrinsic geometry}. The series 
(\ref{z2ext}) will converge for small $t$, hence is a low temperature
expansion (the convergence properties will be discussed in Section 2.3).

It is known that the partition function can also be written in terms 
of loops which have intrinsic geometry:
\begin{equation}
Z \; = \; 2 \; t^{-2V} \sum_{\gamma \in {\cal L}_{int} } (t^2)^{|\gamma|} \;
(-1)^{n(\gamma)} \; .
\label{z2int}
\end{equation} 
Here we introduced the number of self-intersections 
$n(\gamma) = 0,1,2, \dots$ of a loop $\gamma$.
We define ${\cal L}_{int}$ to be the set of all closed, not necessarily
connected, loops $\gamma$ on the dual lattice with a fixed chosen 
orientation and 
with the restriction that each of the links of $\gamma$
is occupied only once. The loops may however intersect themselves
or each other. Thus a loop $\gamma$ in ${\cal L}_{int}$ consists of a 
base point (or several base points when there are disconnected pieces)
and a set of directions with the above restrictions. This is 
what we refer to as loops with {\it intrinsic geometry}.

A representation of the partition function in terms of loops with 
intrinsic geometry is a powerful tool, since it allows to exponentiate
the sum over loops in (\ref{z2int}). This gives an expression of the free 
energy in terms of loops. We will discuss the precise form of this 
representation later in detail. The density of the free energy is a 
well defined physical quantity also in the infinite volume limit,
and its representation in terms of loops is a beautiful illustration 
of the interplay between dynamics and geometry. 

\begin{figure}[htbp]
\epsfysize=0.8in
\hspace*{8mm}
\epsfbox[0 0 442 81] {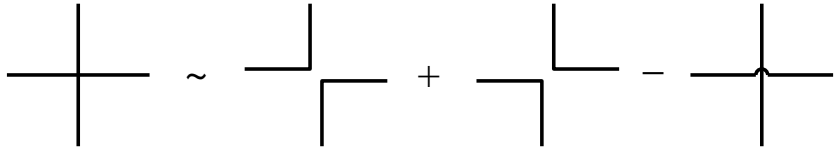}
\caption{{\sl Scheme for the cancellation of overcounting in the
representation} (\protect{\ref{z2int}}) {\sl using loops with intrinsic 
geometry. The left-most 
picture is the pinch-structure for the extrinsic geometry loop, and
the other 3 pictures show the corresponding decomposition in the intrinsic
geometry picture. The last one has an extra minus sign due to the 
intersection factor} $(-1)^{n(\gamma)}$. \label{selfinter}}
\end{figure}

There is a canonical mapping from ${\cal L}_{int}$ onto ${\cal L}_{ext}$,
with the image of a loop $\gamma_{int}$ under this mapping given by the
collection of links traced out by that loop $\gamma_{int}$. Obviously 
there exist many loops in ${\cal L}_{int}$
with distinct intrinsic geometries which get mapped 
to a single loop in ${\cal L}_{ext}$. However, the self-intersection 
factor $(-1)^{n(\gamma)}$ leads to a cancellation of this over-counting
in the sum (\ref{z2int}).
This is illustrated in Fig.~\ref{selfinter}. Every pinch structure of a 
loop in ${\cal L}_{ext}$ (as depicted in Fig.~\ref{selfinter}, left-hand side)
is the image of each of the three pieces of loops in ${\cal L}_{int}$
shown on the right-hand side of the figure.
\vspace{3mm}

Note that larger classes of loops than ${\cal L}_{int}$ can be
used to represent the Ising model. 
For example, the restriction that links 
are occupied only once in a loop can be relaxed to the restriction that the 
loops are non-back-tracking, i.e.~they must not turn around at a site and run 
back on their last link. Again, the same self-intersection factor 
$(-1)^{n(\gamma)}$ gives rise to the necessary cancellations. However, 
for our presentation the above definition of ${\cal L}_{int}$ is 
most convenient.
\\

It is straightforward to generalize (\ref{z2int}) to the case of the 
variable bond Ising model, where the coupling $\beta_{\langle x, y \rangle}$ 
is allowed to vary over links $\langle x, y \rangle$. The generalization of 
(\ref{z2int}) to the case of the variable bond model is given by
\begin{equation}
Z \; = \; 2 \prod_{\langle x, y \rangle} (t_{\langle x, y \rangle})^{-1}
\sum_{\gamma \in {\cal L}_{int}} (-1)^{n(\gamma)}
\prod_{\langle x, y \rangle \in \gamma^*} (t_{\langle x, y \rangle})^2 \; .
\label{z2local}
\end{equation}
Here we have generalized $t$ to  
$t_{\langle x, y \rangle} =$ exp$(-\beta_{\langle x, y \rangle})$ and 
$\gamma^*$ denotes the collection of links on the original lattice
dual to the loop $\gamma$.

All of these representations can be generalized to the 3-$D$ case.
The representation analogous to (\ref{z2ext}) is given by a sum over 
random surfaces with extrinsic geometry,
\begin{equation}
Z \; = \; 2 \; t^{-3V} \sum_{\sigma \in {\cal S}_{ext} } 
(t^2)^{|\sigma|} \; .
\label{z3ext}
\end{equation} 
Here ${\cal S}_{ext}$ denotes all closed, but not necessarily connected
surfaces $\sigma$ made from plaquettes 
on the dual lattice which can be obtained by enclosing lumps of negative
spins in the surface $\sigma$. Analogous to the 2-$D$ case the surface 
$\sigma$ in ${\cal S}_{ext}$ is a collection of plaquettes with zero 
boundary. We introduced the notation 
$|\sigma|$ for the area of the surface $\sigma$.

Also for the 3-$D$ case one is interested in a representation in terms 
of surfaces that allow for intrinsic geometry 
\cite{fradkin},\cite{itzykson2},\cite{sedra}-\cite{distler}. 
It is given by 
\begin{equation}
Z \; = \; 2 \; t^{-3V} \sum_{\sigma \in {\cal S}_{int} } 
(t^2)^{|\sigma|} \; (-1)^{L(\sigma)} \; .
\label{z3int}
\end{equation}
Here $L(\gamma)$ 
denotes the number of links where the surface self-intersects. 
Again different choices for the set ${\cal S}_{int}$ of surfaces 
with intrinsic geometry are possible. We will be more explicit on the
possible choices of ${\cal S}_{int}$ in Section 3, where we also give an elegant 
proof of formula (\ref{z3int}).

\section{The 2-D case}
\setcounter{equation}{0}
In this section we discuss the 2-dimensional case. This serves to outline
the general strategy for the hopping expansion,
to introduce some notation and we also obtain results that
will be needed for the 3-$D$ case. Finally we give the correct
result for the loop expansion of the free energy (exponentiation formula)
which contains an additional factor for iterated loops which has
previously been overlooked in the literature. 

\subsection{Grassmann representation and hopping expansion}
The partition function  in the form 
(\ref{z2ext}) can be written as an integral over 
Grassmann variables \cite{fradkin,samuel1,itzykson1},
which in two dimensions is given by
\begin{equation}
Z = \int\! \prod_{x \in \Lambda^*}
d\eta_{-1}(x) d\eta_{+1}(x) d\eta_{-2}(x) d\eta_{+2}(x) \; 
e^{ \beta [ S_{L} (\eta ) + S_{C} (\eta )
+ S_{M} (\eta ) ]}
\; ,
\label{zferm2d}
\end{equation}
where the line-, corner- and monomer-terms of the action are given by
\begin{eqnarray}
S_{L} ( \eta )\!\!\!& \!=\! &\!\! t^2 \sum_{x \in \Lambda^*}\!\left[ 
\eta_{+1}(x) \eta_{-1}(x+\hat{1}) + \eta_{+2}(x) \eta_{-2}(x+\hat{2}) 
\right] \; ,
\nonumber \\
S_{C} ( \eta )\!\!\!& \!=\! &\!\!\!\!\sum_{x \in \Lambda^*}\!\left[
\eta_{+1}(x) \eta_{-2}(x)\!+\!\eta_{+2}(x) \eta_{-1}(x)\!+\!
\eta_{+2}(x) \eta_{+1}(x)\!+\!\eta_{-2}(x) \eta_{-1}(x) \right],
\nonumber \\
S_{M} (\eta)\!\!\!& \!=\! &\!\!\!\!\sum_{x \in \Lambda^*}\!\left[
\eta_{-1}(x) \eta_{+1}(x) + \eta_{-2}(x) \eta_{+2}(x) \right] \; .
\label{action2d}
\end{eqnarray}
When the exponent in (\ref{zferm2d}) is expanded, 
the non-vanishing contributions to the Grassmann integral exactly reproduce 
(without the overall factor $t^{-2V}$) 
the representation (\ref{z2ext}) of the 
partition function as a sum of loops
with extrinsic geometry. The subscripts $L, C$ and $M$ refer to the 
elements: lines, corners and monomers of the loops as they
are produced by the corresponding terms in the action. These building blocks
give the extrinsic geometry loops in the following way: A line coming 
into a site has to have a partner going out. This property of having 
a partner is enforced by the integration rules for Grassmann variables.
The outgoing line can continue in the direction of the incoming line,
in this case the Grassmann integral is saturated by the monomer term.  
The outgoing line can also turn by $\pi/2$ in which case the 
Grassmann integral is made non-vanishing by the corner terms. The line
can however not turn back since the square of a Grassmann variable vanishes.
Finally there is the possibility that 4 lines are attached to a site and so
saturate the Grassmann integral. This gives rise to the pinch structure 
already discussed above.
Thus the loops produced by the expansion of (\ref{zferm2d}) are closed (every 
incoming line has an outgoing partner) and non-back-tracking (square 
of the Grassmann variable vanishes). They are loops in extrinsic
geometry since they simply occur as a set of links. Thus the expansion of
(\ref{zferm2d}) reproduces (\ref{z2ext}). For details concerning the ordering 
of the Grassmann variables see \cite{samuel1,itzykson1}.

For the following we need to write the action in a more compact and 
also anti-symmetrized form. We introduce the vector
\begin{equation}
\eta(x) \; = \; 
\Big( \eta_{+1}(x), \eta_{-1}(x), \eta_{+2}(x), \eta_{-2}(x) \Big)^T \; ,
\end{equation}
and the $4\times 4$ matrices $P_\mu(i,j)$ (the same matrices will be used
in the 3-$D$ case where it is more convenient to denote them as
$P_{\pm x}$ and $P_{\pm y})$
\begin{eqnarray}
P_{+1} (i,j) \; \equiv \;
P_{+x} (i,j) \; \equiv \; \delta_{i,1} \; \delta_{j,2} & \; , \; & 
P_{-1} (i,j) \; \equiv \;
P_{-x} (i,j) \; \equiv \; - \delta_{i,2} \; \delta_{j,1} \; , 
\nonumber \\ 
P_{+2} (i,j) \; \equiv \; 
P_{+y} (i,j) \; \equiv \;\delta_{i,3} \; \delta_{j,4} & \; , \; & 
P_{-2} (i,j) \; \equiv \;
P_{-y} (i,j) \; \equiv \; - \delta_{i,4} \; \delta_{j,3} \; . 
\nonumber \\
\label{pmatrix}
\end{eqnarray}
They obey $P_\mu^T = - P_{-\mu}$. We also define
\begin{equation}
 M = 
\left( \begin{array}{cccc} 
 0 & -1 & -1 & +1 \\ 
+1 &  0 & -1 & -1 \\ 
+1 & +1 &  0 & -1 \\ 
-1 & +1 & +1 & 0 
\end{array} \right) 
\; \; \; \; \; \mbox{with} \; \; \; \; \; 
M^{-1} = 
\left( \begin{array}{cccc} 
 0 & -1 & +1 & -1 \\ 
+1 &  0 & +1 & +1 \\ 
-1 & -1 &  0 & -1 \\ 
+1 & -1 & +1 & 0 
\end{array} \right) \; .
\label{mmatrix}
\end{equation}
We remark that $\det M = 1$ and $M^T = -M$. 
With these definitions the action can be written as 
(the overall factor 1/2 comes from the anti-symme\-tri\-zat\-ion)
\begin{equation}
S(\eta) \; = \; \frac{1}{2} \sum_{x,y \in \Lambda}
\; \eta^T(x) \; K(x,y) \; \eta(y) \; \equiv \; \frac{1}{2} \eta^T K \eta \; ,
\label{quadra}
\end{equation}
where the kernel $K$ is given by
\begin{equation}
K(x,y) \; = \; M \delta (x,y) +  
t^2\! \sum_{\mu = \pm 1}^{\pm 2} P_\mu \; \delta(x + \hat{\mu}, y) \; 
\equiv \; M \delta(x,y) + R(x,y) \; .  
\label{kmatrix}
\end{equation}
It is easy to see that $K$ is anti-symmetric. 
We remark that the representation (\ref{quadra}), (\ref{kmatrix}) is a natural
way of writing the action for the Grassmann variables. It was already 
introduced in \cite{dotsenko}. The partition function 
is given by a Pfaffian which, since $K$ is anti-symmetric, reduces to
the root of a determinant\footnote{This holds only for even-dimensional
matrices. As already discussed, 
it is possible to work on a finite lattice with open 
boundary conditions and perform the infinite volume limit in the end. Since
we have 4 components of Grassmann variables $K$ is always even-dimensional.}
\begin{equation}
Z = \int d \eta \; e^{\frac{1}{2} \eta^T K \eta} = \; 
\mbox{Pf} \; K  =  \sqrt{ \det K} \; = \; \sqrt{\det[ M + R ]} =  
\sqrt{\det [ 1 + M^{-1} R ]} .
\label{zferm}
\end{equation}
Expanding the determinant one obtains
\begin{equation}
Z \;  = \; 
\exp \left(- \frac{1}{2} \sum_{n=1}^\infty \frac{ (-t^2)^n}{n} 
\mbox{Tr} \; [H^n] \right) ,
\label{hopexp}
\end{equation}
where the hopping matrix $H$ is defined as $t^2H \equiv M^{-1} R$. In the
last step we used the well known formula for the expansion of 
determinants of the form $\det[1- t^2 H]$. This series converges for 
$t^2 \parallel \! H \! \parallel_{\infty} < 1$, and in Section 2.3 we will show
that $t = t_{crit} \equiv \exp(-\beta_{crit})$ saturates this bound.
The hopping matrix $H$ has the form 
\begin{equation}
H(x,y) \; = \; \sum_{\mu = \pm 1}^{\pm 2} H_\mu \; \delta(x + \hat{\mu}, y) 
\; \; \; \; \mbox{with} \; \; \;  H_\mu \equiv M^{-1} P_\mu \; .
\label{hopmat}
\end{equation}
The trace of powers of $H$ is given by 
\begin{eqnarray}
\mbox{Tr}[ H^n ]\!\!\!& = &\!\!\sum_{x_1, x_2 \dots x_n}
\sum_{\mu_1, \mu_2 \dots \mu_n}\!\!
\delta(x_1 + \hat{\mu}_1, x_2) \dots
\delta(x_n + \hat{\mu}_n, x_1) \; \mbox{Tr}[ H_{\mu_1} 
\dots H_{\mu_n} ] 
\nonumber \\
& = & \sum_{x_1} \sum_{\mu_1, \mu_2 \dots \mu_n} 
\delta(x_1 +\hat{\mu}_1 + \hat{\mu}_2 + \dots \hat{\mu}_n , x_1) \;  
\mbox{Tr} \; [ H_{\mu_1} H_{\mu_2} \dots H_{\mu_n} ] \; . 
\nonumber \\
\label{htrace}
\end{eqnarray}
Structures of this type are well known from the hopping expansion 
(see e.g.~\cite{MoMu94}) of 
the fermion determinant in lattice gauge theories with fermions. Due to the 
Kronecker delta the terms in the sum have support on closed loops. 
In the next section we will show 
that these loops have simple properties due to the algebra of the $H_\mu$'s and
this will also lead to a straightforward computation of the trace 
$\mbox{Tr} \; [ H_{\mu_1} H_{\mu_2} \dots H_{\mu_n} ]$ for arbitrary 
closed loops $\gamma$.

\subsection{Loop representation from the hopping expansion}
\begin{figure}[htbp]
\epsfysize=1.0in
\hspace*{9mm}
\epsfbox[0 0 410 97] {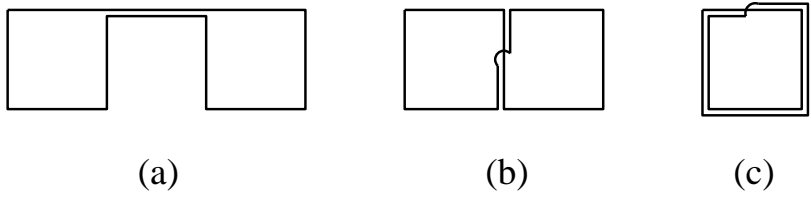}
\caption{ {\sl Some of the loops occuring in the hopping expansion. 
They differ from the loops in ${\cal L}_{int}$ since they can occupy links
several times or iterate their whole path.}
\label{loopsfig}}
\end{figure}

From the remarks in the end of the last section it is clear 
that the loops supporting the contributions to (\ref{htrace}) 
are closed loops with a base point $x_1$
and described by the {\it ordered} set of 
directions $\{ \mu_1, \mu_2, \dots \mu_n \}$, where each 
$\mu_i$ can have the values $\pm 1, \pm 2$. These loops are more general 
than the ones we included in ${\cal L}_{int}$. They are connected and are allowed 
to self-intersect, however in addition, they can occupy links 
several times or even iterate their whole path many times.  
Since closed loops have even length: 
Tr$[H^{2n + 1}] = 0$, and only even terms contribute in (\ref{hopexp}). 
Some examples of loops occuring in
the hopping expansion are depicted in Fig.~\ref{loopsfig}.
\vspace{3mm}

In order to compute the 
weights for the loops in (\ref{htrace}) the properties of the matrices 
$H_\mu = M^{-1} P_\mu$ have to be studied. The $H_\mu$-matrices will
be encountered again in the 3-$D$ case where it will be convenient 
to denote them as $H_{\pm x}$ and $H_{\pm y}$. They are explicitly 
given by (compare \cite{dotsenko})
\begin{eqnarray}
H_{+1} \equiv H_{+x} =  \left( \begin{array}{cccc} 
 0 &  0 &  0 &  0 \\ 
 0 & +1 &  0 &  0 \\ 
 0 & -1 &  0 &  0 \\ 
 0 & +1 &  0 &  0 
\end{array} \right)\!\!\!\!\!\!& , &\! 
H_{-1} \equiv H_{-x} =  \left( \begin{array}{cccc} 
+1 &  0 &  0 &  0 \\ 
 0 &  0 &  0 &  0 \\ 
+1 &  0 &  0 &  0 \\ 
+1 &  0 &  0 &  0 
\end{array} \right) ,
\nonumber \\
H_{+2} \equiv H_{+y} =  \left( \begin{array}{cccc} 
 0 &  0 &  0 & +1 \\ 
 0 &  0 &  0 & +1 \\ 
 0 &  0 &  0 &  0 \\ 
 0 &  0 &  0 & +1 
\end{array} \right)\!\!\!\!\!\!& , &\!
H_{-2} \equiv H_{-y} =  \left( \begin{array}{cccc} 
 0 &  0 & +1 &  0 \\ 
 0 &  0 & -1 &  0 \\ 
 0 &  0 & +1 &  0 \\ 
 0 &  0 &  0 &  0 
\end{array} \right) . \nonumber \\
\label{hmatrices}
\end{eqnarray}
We will now show that the algebra of these matrices restricts the set of 
loops occuring in the hopping expansion.
The first observation is that $H_{\pm \mu} H_{\mp \mu} = 0$,  
this property excludes back-tracking loops in (\ref{htrace}).
Furthermore one can show that the trace of a product of $H_\mu$'s is 
invariant under reversing the orientation of the loop. This can be seen 
by using the definition $H_\mu = M^{-1} P_\mu$ and the 
transposition properties  
$P_\mu^T = -P_{-\mu}$ and $(M^{-1})^T = - M^{-1}$,
\begin{eqnarray}
\mbox{Tr} \; [ H_{-\mu_{2n}} H_{-\mu_{2n-1}} \dots H_{-\mu_1} ] & = &
\mbox{Tr} \; [ M^{-1} P_{\mu_{2n}}^T \dots M^{-1} P_{\mu_1}^T ] \; = 
\nonumber \\
\mbox{Tr} \; [ P_{\mu_1} M^{-1} \dots P_{\mu_{2n}} M^{-1} ]^T & = & 
\mbox{Tr} \; [ H_{\mu_1} H_{\mu_2} \dots H_{\mu_{2n}} ] \; .
\nonumber
\end{eqnarray}
One can then fix an 
orientation for each loop in (\ref{htrace}) and write a factor 2 in front
of the sum. This factor cancels the overall factor of 1/2 from the square 
root in  (\ref{zferm}).

\begin{figure}[htbp]
\epsfysize=1.7in
\hspace*{20mm}
\epsfbox[0 0 294 147] {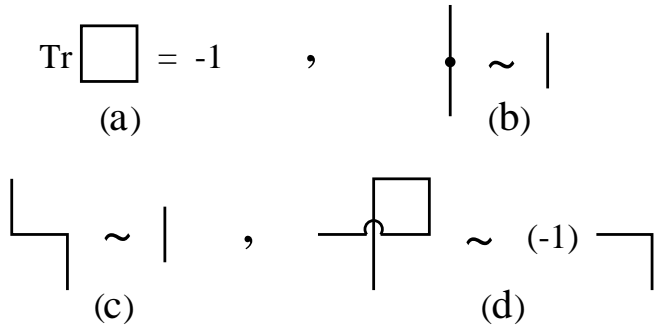}
\caption{ {\sl Graphical representation of the rules for the reduction of
$\mbox{Tr} \; [ \prod_{\mu \in \gamma} H_\mu]$: 
basic loop (a), telescope rule (b),
kink rule (c) and intersection rule (d) }.
\label{rules}}
\end{figure}

Thus far we have shown that the paths which contribute in (\ref{htrace}) are 
closed, connected, non back-tracking loops, $\gamma$, with a chosen 
orientation. We finally prove that the trace over the product of $H_\mu$'s along 
a loop $\gamma$ yields the self-intersection factor
\begin{equation}
\mbox{Tr} \; \prod_{\mu \in \gamma} H_\mu \; = 
\; - (-1)^{n(\gamma)} \; .
\label{traceformula}
\end{equation}
As before $n(\gamma)$ denotes the number of self-intersections of the loop $\gamma$.
The proof is decomposed into several steps. Using the explicit
form (\ref{hmatrices}) of the $H_\mu$'s it is straightforward to show the following rules:
\\
\\
{\it Basic loop}:
\begin{equation} 
\mbox{Tr} [ H_{+1} H_{+2} H_{-1} H_{-2} ] \; = \; -1 \; .
\label{basic}
\end{equation}
{\it Telescope rule}:
\begin{equation}
H_\mu^2 \; = \; H_\mu \; .
\label{telescope}
\end{equation}
{\it Kink rule}:
\begin{equation}
H_\mu H_\nu H_\mu \; = \; H_\mu \; \; \; \mbox{for} \; \; \; \nu \neq - \mu \; .
\label{kink}
\end{equation}
{\it Intersection rule}:
\begin{eqnarray}
(H_{+1})^2 H_{+2} H_{-1} (H_{-2})^2  & = & - H_{+1} H_{-2} \; , 
\nonumber \\
(H_{+2})^2 H_{+1} H_{-2} (H_{-1})^2  & = & - H_{+2} H_{-1} \; .
\label{intersection}
\end{eqnarray}
The four rules have the obvious graphical interpretation depicted 
in Fig.~\ref{rules}. 

The result (\ref{basic}) is just formula (\ref{traceformula})
for the simplest loop, i.e.~the one running around a single plaquette. 
Formula (\ref{telescope}) allows one to stretch or shrink a
loop without altering the trace (\ref{traceformula}). The kink rule 
(\ref{kink}) allows one to remove kinks in the path, and finally the 
intersection rule (\ref{intersection})
allows one to remove sub-loops, giving rise to a factor of $-1$ for each 
self-intersection. Formula (\ref{intersection}) gives the result 
(for both possible orientations) for the sub-loop depicted in Fig.~\ref{rules}~(d). 
The other 6 possibilities for (\ref{intersection})
follow immediately from invariance of the formalism
under rotations by $\pi/2$.

\begin{figure}[htbp]
\epsfysize=2.0in
\hspace*{18mm}
\epsfbox[0 0 344 194] {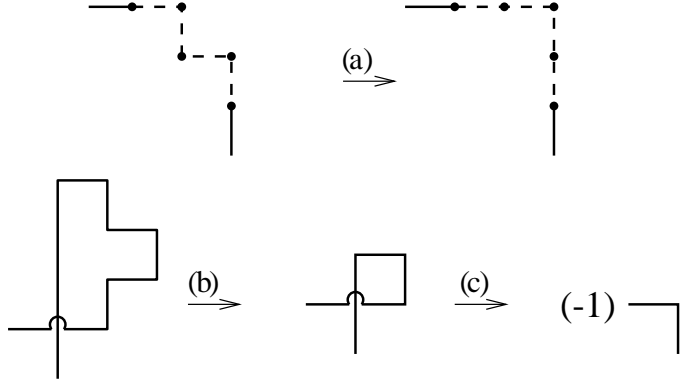}
\caption{
{\sl Examples for the reduction of loops: In step (a) telescope and
kink rule were used to smoothen the dotted part of the loop
(compare} (\protect{\ref{subchain}}) {\sl  for the corresponding algebraic
expression). In step (b) we bring a sub-loop to canonical form,
again using telescope and kink rule, and in step (c)
the sub-loop is removed using the intersection rule giving an overall 
factor $-1$.}
\label{redex}}
\end{figure}

The above rules allow for a constructive reduction of the trace 
Tr$\prod_{\mu \in \gamma} H_\mu$ for an arbitrary closed, connected,
non-back-tracking loop $\gamma$ as follows:
\\
\\
(1) Start with a sub-loop which has only one self-intersection point  
(an example for such a sub-loop is e.g.~given in 
Fig.~\ref{redex}, second line). If the loop we started with 
has no self-intersection at all, proceed to (2), otherwise there exist
at least two such sub-loops.
\\
\\
(2) Use telescope and kink rules to bring the sub-loop to the standard form 
as depicted in the left picture of 
Fig.~\ref{rules}~(d). Under these transformations the
trace (\ref{traceformula}) remains invariant.
In case the loop had no self-intersection 
to begin with, bring it to the standard form for loops without self-intersection
as depicted in Fig.~\ref{rules}~(a), and (\ref{basic}) is the final result.
\\
\\
(3) Use the intersection rule to remove the sub-loop and collect an
overall factor of $-1$. Repeat the steps (1) - (3) until finished.
\\
\\
We remark that if nested sub-loops coincide on some links, it is 
always possible to disentangle them using the telescope rule.
Let's elaborate a little bit more on the actual implementation of step (2):  
The idea is to replace products of $H_\mu$'s, corresponding to 
some sub-chain of links, by another product of $H_\mu$'s (obtained 
from the first one using kink and telescope rules), such that the 
new chain corresponds to a new piece of loop which is smoother,
but still leaves the whole loop closed. The necessary condition for the
new sub-chain to match the starting- and end-point of the old sub-chain 
is that the number of $+1$-moves minus the number of 
$-1$-moves as well as the number of $+2$-moves minus the
number of $-2$-moves remains invariant (compare the example in 
Fig.~\ref{redex}~(a).
In order to illustrate the steps involved we discuss two examples
(see Fig.~\ref{redex}).
\vspace{3mm}

The first two pictures in Fig.~\ref{redex}
demonstrate how the telescope and kink rules
are used to smoothen a contour (step (a) in the figure).
The dotted part of the contour in the left figure 
is represented by the left-hand side of the following equation 
\begin{equation}
H_{+1} H_{-2} H_{+1} H_{-2} \; = \; H_{+1} H_{-2} \; = \; 
(H_{+1})^2 (H_{-2})^2 \; .
\label{subchain}
\end{equation}
Here we used the kink rule (\ref{kink}) to replace 
$H_{+1} H_{-2} H_{+1}$ by $H_{+1}$ and then expanded, using the telescope rule 
(\ref{telescope}), to obtain the form on the right-hand side. This is once again
a contour which matches starting- and end-points and corresponds to the 
second dotted contour in Fig.~\ref{redex}~(a). 
Thus step (a) in Fig.~\ref{redex} is simply the replacement of the 
left-hand side of (\ref{subchain}) inside some
trace over $H_\mu$'s by its right-hand side. 
With combinations of telescope and kink rules all
(sub-) loops can be transformed into squares, which can then be
shrunk to loops around single plaquettes using the telescope rule. 
Under such a set of operations the
trace (\ref{traceformula}) remains invariant. An
example of such a transformation is depicted in step (b) of Fig.~\ref{redex}.
Finally in step (c) of Fig.~\ref{redex} the sub-loop
is removed using the intersection rule (\ref{intersection}) 
and an overall factor of $-1$ emerges.
These steps can be applied iteratively to prove (\ref{traceformula}) for
an arbitrary closed, connected, non-back-tracking loop $\gamma$. 

On inserting (\ref{traceformula}) in (\ref{htrace}) and the result in 
(\ref{hopexp}) one finds
\begin{equation}
Z \; = \; \mbox{Pf} K
\; = \; \exp \left( \sum_{n=1}^\infty \frac{ (t^2)^{2n}}{2n}
\sum_x \sum_{\gamma \in L^{(2n)}_x} (-1)^{n(\gamma)} \right) \; .
\label{expsum1}
\end{equation}
Here $L_x^{(2n)}$ denotes the set of closed, connected,
non-back-tracking loops of length $2n$ based at $x$ with a chosen orientation.
Recall that (\ref{htrace}) is non-vanishing only for even $n$ and 
thus the alternating sign of (\ref{hopexp}) disappears.
The overall factor of 1/2 in (\ref{hopexp}) is gone since we chose only one
of the two possible orientations of a loop, and the 
overall minus sign in (\ref{hopexp}) is cancelled by the overall sign in 
(\ref{traceformula}).

It is possible to simplify the exponential in (\ref{expsum1})
even further: To do that we discuss (in order of increasing complexity)
loops with singly occupied links, iterated loops where the whole set of 
links is run through several times and finally loops where only some links 
are occupied several times. To start take some loop $\gamma$
in $L_x^{(2n)}$ where each link in the loop is occupied only once. 
For the contour corresponding to this loop $\gamma$ there are all together 
$2n$ inequivalent choices of a base point. Each of these points can serve as 
the base point for a different loop giving rise to the same contribution 
to (\ref{expsum1}) as $\gamma$. As such,
one can choose a single representative of this class of loops and remove the
factor of $1/2n$ in (\ref{expsum1}). Now consider a loop which runs through 
its contour twice, i.e.~each link occurs exactly twice (an example of such a 
loop is given in Fig.~\ref{loopsfig}~(c)). 
For such an iterated loop we have only $2n/2$ possible choices of inequivalent
base points. In general for a loop $\gamma$ that is iterated $I(\gamma)$-times
there are $2n / I(\gamma)$ inequivalent choices for the base point. As for the 
non-iterated loops, one can choose a single representative and remove the factor
of $1/2n$ in (\ref{expsum1}), however, the factor of $1/I(\gamma)$ remains.
Finally we remark that for loops where
only a sub-loop is iterated no such factor can occur, since the set of
directions $\mu_1, \mu_2 \dots \mu_{2n}$ is different
(cyclicly permuted) for some starting 
point $x$, and the same point visited by the loop after running through 
a sub-loop. The partition function thus reads:
\begin{equation}
Z \; = \; \mbox{Pf} K
\; = \; \exp \left( \sum_{\gamma \in {\cal L}^{con}_{exp}} 
(t^2)^{|\gamma|} \; (-1)^{n(\gamma)} \; \frac{1}{I(\gamma)} \right) \; ,
\label{expsum2}
\end{equation}
where ${\cal L}_{exp}^{con}$ denotes the set of all closed, connected,
non-back-tracking loops (without base point) 
and $I(\gamma)$ is the number of iterations of the
loop $\gamma$.

\begin{figure}[htbp]
\epsfysize=0.6in
\hspace*{22mm}
\epsfbox[0 0 321 60] {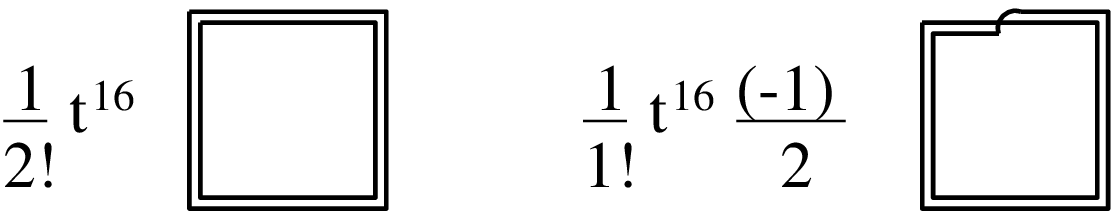}
\caption{
{\sl Example for the cancellation of an iterated loop. The figure on the 
left-hand side shows the contribution of the basic loop around a single
plaquette in the quadratic term of the expansion of the exponential 
in }(\protect{\ref{expsum2}}). {\sl Here the factor 1/2 comes from the 
power series expansion of the exponential function. The picture on the 
right-hand side shows the contribution of the iterated loop, which appears
in the linear term of the expansion. Here the factor 1/2 comes from the 
iteration term $1/I(\gamma)$. The self intersection term produces the
minus sign which ensures the cancellation of the two contributions.}
\label{example}}
\end{figure}

We stress that this result differs from previous formulas 
for the exponentiation of (\ref{z2int}) by the factor 
$1/I(\gamma)$ for iterated loops. This factor is, however, essential
for the correct cancellation of loops that have no counterpart in 
${\cal L}_{ext}$. This can be seen by the 
following simple example: Consider the basic loop of length 4 which runs around
a plaquette. When the exponent in (\ref{expsum2}) is expanded, this loop 
will, in the quadratic term of the expansion, give rise to the plaquette
which is occupied by two such loops running around this single plaquette
(see the left-hand side of Fig.~\ref{example}).
It has the 
factor $t^{16} / 2$, where the factor 1/2 comes from the expansion of
the exponential function. This loop has no counterpart in ${\cal L}_{ext}$
and thus has to be cancelled by some other loop. It is clear that the only 
loop which can cancel this loop is the one depicted on the right-hand 
side of Fig.~\ref{example}.
It appears in the linear term of the expansion of the exponential function
and comes with the factor $- t^{16} / 2$ and thus exactly cancels the first
loop. The overall minus sign is due to the self-intersection and the factor 
1/2 comes from the term $I(\gamma)^{-1}$. 
Hence this factor is essential for the cancellation of unphysical loops.
\vspace{3mm}

With the derivation of (\ref{expsum2}) we have established that the 
hopping expansion (\ref{hopexp}) of the Grassmann formulation gives an
exact algebraic representation of the loop expansion.

It is straightforward to generalize (\ref{expsum2}) to the case of the 
variable bond Ising model by running through the derivation once more
but allowing for varying Boltzmann factors $t_{\langle x, y \rangle} =$ 
exp$(-\beta_{\langle x, y \rangle})$.
The result is (neglecting the overall factor 
$2 \prod (t_{\langle x, y \rangle})^{-1}$):
\begin{equation}
Z = \exp \left( \sum_{\gamma \in {\cal L}^{con}_{exp}} 
\frac{(-1)^{n(\gamma)}}{I(\gamma)} \; 
\prod_{\langle x, y \rangle \in \gamma^*}
(t_{\langle x, y \rangle})^2 \right) \; .
\label{expsumlocal}
\end{equation}
We remark that when a loop $\gamma$ occupies some links several times, 
its dual $\gamma^*$ has multiply occupied links. Thus in the last equation 
the product includes a link-factor whenever the link is intersected by 
the loop. In particular for iterated loops each link factor 
$(t_{\langle x, y \rangle})^2$ occurs $I(\gamma)$-times.

\subsection{Radius of convergence and critical temperature}
The formula (\ref{expsum2}) is a physically interesting result, since
the exponent is proportional to the
loop expansion of the free energy density $f$
\begin{equation}
f \; = \; - \frac{1}{\beta} \lim_{V \rightarrow \infty} 
\frac{1}{V} \sum_{\gamma \in {\cal L}^{con}_{exp}} 
(t^2)^{|\gamma|} \; (-1)^{n(\gamma)} \; \frac{1}{I(\gamma)} \; .
\label{freeE}
\end{equation}
The sum (\ref{freeE}) will converge for sufficiently small $t$, and the
radius of convergence 
corresponds to the critical temperature. However, computing the radius of 
convergence of the loop expansion in the form (\ref{freeE}) is a rather 
intractable problem. It is very convenient to make use of the form 
(\ref{hopexp}) of the expansion, where it is clear that
the expansion converges for 
\begin{equation}
t^2 \parallel \!\! H \!\! \parallel_{\infty} \; < \; 1 \; .
\label{converge}
\end{equation}

The norm $\parallel\!H\! \parallel_\infty$ is defined as 
sup$_{\parallel t \parallel = 1} \sqrt{ (t,H^\dagger H t)}$ where 
$t$ is some test-function  and $\parallel\! t \! \parallel = \sqrt{(t,t)}$,
and the inner product is defined to be the $l^2$ product obtained 
by summing over all lattice and spinor indices. It is straightforward 
to compute
\[
H^\dagger H \!=\! \left(\!\!\!\begin{array}{cccc} 
3 \delta(x,y) &  0 & \!\!\!\!\!
2 \delta(x\!+\!\hat{1}\!-\!\hat{2},y) &  
2 \delta(x\!+\!\hat{1}\!+\!\hat{2},y)  \\ 
0 & 3 \delta(x,y)  & \!\!\!\!\!
-2 \delta(x\!-\!\hat{1}\!-\!\hat{2},y) & 
2 \delta(x\!-\!\hat{1}\!+\!\hat{2},y)  \\ 
2 \delta(x\!-\!\hat{1}\!+\!\hat{2},y) & 
-2 \delta(x\!+\!\hat{1}\!+\!\hat{2},y) &  \!\!\!\!\! 
3 \delta(x,y) &  0 \\ 
2 \delta(x\!-\!\hat{1}\!-\!\hat{2},y) & 
2 \delta(x\!+\!\hat{1}\!-\!\hat{2},y) &  \!\!\!\!\!
0 &  3 \delta(x,y)  
\end{array}\!\!\!\right)\!. 
\]
The lattice indices of this matrix can be diagonalized using 
Fourier transformation, giving
\[
F^\dagger H^\dagger H F (p,q) = 
\delta(p,q)\!\left(\!\!\!\begin{array}{cccc} 
3 & \! 0 & \!\!\!\!
2 e^{-ip_1 + ip_2} &  2 e^{-ip_1 - ip_2}   \\ 
0 & \! 3 & \!\!\!\! 
-2 e^{+ip_1 + ip_2}  & 2 e^{+ip_1 - ip_2} \\ 
2 e^{+ip_1 - ip_2} & \! -2 e^{-ip_1 - ip_2} &  \!\!\!\!
3 &  0 \\ 
2 e^{+ip_1 + ip_2} &\! 2 e^{-ip_1 + ip_2} & \!\!\!\! 
0 &  3   
\end{array}\!\!\! \right)\!,
\]
where the transformation matrix $F$ is given by $F(y,q) = \exp(-iyq)/2\pi$.
The eigenvalues $\lambda$ for the remaining $4\times 4$ problem can be 
easily computed giving (each $\lambda$ is two-fold degenerate)
\begin{equation}
\lambda_\pm \; = \; 3 \pm 2 \sqrt{2} \; .
\label{eigenvalues}
\end{equation}
The norm $\parallel \! H \! \parallel_\infty$ is then given by the square root
of the larger eigenvalue. From (\ref{converge}) and (\ref{eigenvalues})
we obtain for the
critical $t_c = e^{-\beta_c}$:
\begin{equation}
1 \; = \; (t_c)^2 \parallel \! H \! \parallel \; = \; e^{-2 \beta_c} 
\sqrt{3 +2 \sqrt{2}} \; \; \;  \Longrightarrow \; \; \; \beta_c \; = \; 
\frac{1}{2} \ln( 1 + \sqrt{2}) \; .
\end{equation}
This is the well known result for the critical inverse temperature.
We thus have achieved an elegant proof that the radius of convergence of the 
loop expansion (\ref{freeE}) for the free energy 
is determined by the critical temperature. 
Computing the radius of convergence of (\ref{freeE}) without the Grassmann 
representation is a rather intractable problem.

\section{The 3-D case}
\setcounter{equation}{0}

We now discuss the 3-$D$ case. We will see that the 3-$D$ model factorizes 
into products of 2-$D$ models coupled to an auxiliary field
and many of the concepts developed in the previous
section will become very useful. 

\subsection{Grassmann representation and hopping expansion}
Also for the 3-$D$ case there exists a Grassmann integral for the 
representation of the partition function in terms of surfaces with
extrinsic geometry \cite{fradkin,samuel3,itzykson2}. Here we use a 
notation different from 
the original work, which is however more convenient for our expansion. 

\begin{figure}[htbp]
\epsfysize=2.5in
\hspace*{25mm}
\epsfbox[155 155 440 375] {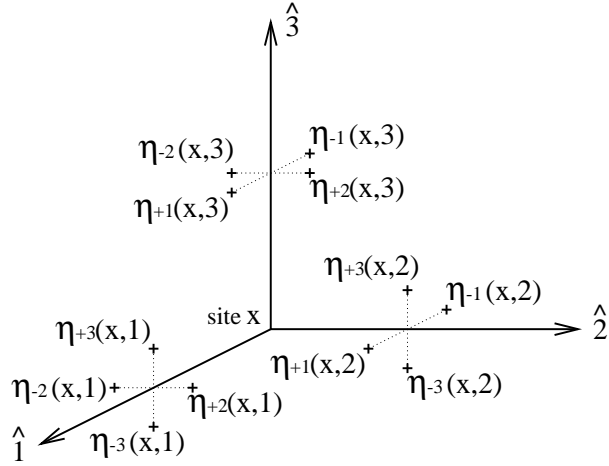}
\caption{{\sl Labeling of the Grassmann variables $\eta_{\pm \nu}(x,\mu)$ 
associated with the links of the dual lattice.}
\label{3dgrass}}
\end{figure}

Each link of the dual lattice carries 4 Grassmann variables denoted by
$\eta_{\pm \nu} (x, \mu )$ as depicted in Fig.~\ref{3dgrass}. 
Here $x$ and $\mu$ denote the link the variable
lives on, and $\pm \nu$  distinguishes the 4 different
variables on each link.

The action for the Grassmann variables is again a sum of three terms
$S = S_P + S_E + S_M$
producing the elements of the surfaces: plaquettes, edges and monomers.
The terms are given by
\begin{eqnarray}
S_P[\eta] & = & t^2 \sum_x \sum_{\stackrel{\scriptstyle \mu, \nu = 1}{\mu < \nu}}^3
\eta_{+\mu}(x,\nu) \eta_{-\mu}(x+\hat{\mu},\nu) \;
\eta_{+\nu}(x,\mu) \eta_{-\nu}(x+\hat{\nu},\mu) \; , 
\nonumber \\
S_E[\eta] & = & \sum_x \sum_{\mu = 1}^3 
\sum_{\stackrel{\scriptstyle \nu, \rho \neq \mu}{\nu < \rho}} \Big[ 
\eta_{+\nu}(x,\mu) \eta_{-\rho}(x,\mu) +
\eta_{+\rho}(x,\mu) \eta_{+\nu}(x,\mu)  
\nonumber \\
 & + & \eta_{+\rho}(x,\mu) \eta_{-\nu}(x,\mu) +
\eta_{-\rho}(x,\mu) \eta_{-\nu}(x,\mu) \Big] \; ,
\nonumber \\
S_M[\eta] & = & \sum_x \sum_{\mu = 1}^3 
\sum_{\stackrel{\scriptstyle \nu, \rho \neq \mu}{\nu < \rho}} \Big[ 
\eta_{-\nu}(x,\mu) \eta_{+\nu}(x,\mu) +
\eta_{-\rho}(x,\mu) \eta_{+\rho}(x,\mu) \Big] \; .
\label{3daction}
\end{eqnarray}
The Grassmann representation of the 3-$D$ Ising model 
\cite{fradkin,samuel3,itzykson2}
is a straightforward generalization of the 2-$D$ case. 
Again the exponential of (\ref{3daction})
has to be expanded, and the Grassmann integral reproduces the 
terms corresponding to the representation of the partition function in 
terms of surfaces with extrinsic geometry. Similar to the 2-$D$ case 
plaquettes, produced by $S_P$, that are incoming to a link must have an
outgoing partner to saturate the Grassmann integral. Possible choices for
the direction of the outgoing plaquette are either the same direction as
the incoming plaquette (monomer terms) or rotated by $\pi/2$ (edge terms).
The outgoing plaquette can however not fold back on the incoming one
due to the fact that the square of a Grassmann variable vanishes. Also
4 plaquettes attached to a link can saturate the Grassmann integral, 
producing the pinch structure of the surfaces. The 
resulting terms of the Grassmann integral thus correspond to the closed surfaces
in ${\cal S}_{ext}$. For details concerning the ordering 
of the Grassmann variables see \cite{fradkin,samuel3,itzykson2}.

The quartic interaction, from $S_P[\eta]$, can be decomposed into a Yukawa-like term 
with a tensor field by using the following identity for
Grassmann variables $\eta_1, \eta_2, \eta_3, \eta_4$
\begin{equation}
\exp \left( t^2 \eta_1 \eta_2\eta_3\eta_4 \right)  = \frac 1 2 \sum_{A =\pm 1} 
\exp \Big( t~A~[ \eta_1 \eta_2 + \eta_3\eta_4 ] \Big) \; .
\end{equation}
Thus, $S_P[\eta]$ can be replaced by
\begin{equation}
S_P[\eta, A] = t \sum_x \sum_{\stackrel{\scriptstyle \mu, \nu = 1}{\mu < \nu}}^3
A_{\mu \nu}(x) \Big[
\eta_{+\mu}(x,\nu) \eta_{-\mu}(x+\hat{\mu},\nu) +
\eta_{+\nu}(x,\mu) \eta_{-\nu}(x+\hat{\nu},\mu) \Big], 
\end{equation}
and the partition function now contains a sum over auxiliary fields
$A_{\mu \nu}(x)$ associated with the plaquettes $(x; \mu, \nu)$ of the
dual lattice,
\begin{equation}
Z \; = \; \left( \frac{1}{2} \right)^{3 V} 
\sum_{ \{A_{\mu \nu}(x) = \pm 1 \}} \int d\eta \; e^{S_P[\eta, A] +
S_E[\eta] + S_M[\eta] } \; .
\end{equation}
The introduction of the auxiliary fields transformed the action for the 
Grassmann variables to a quadratic form, and the hopping expansion 
methods can be  applied.
As in the 2-$D$ case, the next step is to introduce the vectors $\psi(x)$ containing all 
12 variables living on the 3 links associated with the site $x$ 
of the dual lattice
\begin{eqnarray}
\psi(x) & = & \Big( 
\eta_{+2}(x,1), \eta_{-2}(x,1), \eta_{+3}(x,1), \eta_{-3}(x,1), 
\eta_{+1}(x,2), \eta_{-1}(x,2), 
\nonumber \\
& & \eta_{+3}(x,2), \eta_{-3}(x,2),
\eta_{+1}(x,3), \eta_{-1}(x,3), \eta_{+2}(x,3), \eta_{-2}(x,3) \Big)^T \; .
\nonumber
\end{eqnarray}
With this ordering of the variables, the edge and monomer terms of
the action can be written as a quadratic form 
$S_E + S_M = 1/2 \psi^T \widetilde{M} \; \psi$,
which is a simple generalization of the expression already used in the
2-$D$ case. The kernel $\widetilde{M}$ for the corresponding terms is 
block-diagonal, with the blocks given by the matrix $M$ from the 2-$D$
formulation (\ref{mmatrix}),
\begin{equation}
\widetilde{M} \; \equiv \; \mbox{diag} \; \Big( 1, 1, 1 \Big) \otimes M\; .
\end{equation}
$\widetilde{M}$ inherits the following properties from $M$:
det $\widetilde{M} = 1, \widetilde{M}^T = -\widetilde{M}$ 
and $\widetilde{M}^{-1} =$ 
diag$(1, 1, 1)\otimes M^{-1}$ with $M^{-1}$ given in (\ref{mmatrix}).
Also, the terms containing the auxiliary fields have a simple form if one makes 
use of the matrices $P_{\pm x}, P_{\pm y}$ already 
defined in (\ref{pmatrix}),
\begin{equation}
S_P \; = \; \frac{t}{2} \sum_x \sum_{\mu = \pm 1}^{\pm 3} 
\psi(x)^T B_\mu(x) \psi(x + \hat{\mu}) \; ,
\end{equation}
where the fields $B_\mu$ are composed from the auxiliary fields 
$A_{\mu \nu} (x)$ as follows:
\begin{eqnarray}
B_{+1} (x) & = & A_{12}(x) \; \mbox{diag}(0,1,0) \otimes P_{+x} +
A_{13}(x) \; \mbox{diag}(0,0,1) \otimes P_{+x} \; ,
\nonumber \\
B_{+2} (x) & = & A_{12}(x) \; \mbox{diag}(1,0,0) \otimes P_{+x} +
A_{23}(x) \; \mbox{diag}(0,0,1) \otimes P_{+y} \; ,
\nonumber \\
B_{+3} (x) & = & A_{13}(x) \; \mbox{diag}(1,0,0) \otimes P_{+y} +
A_{23}(x) \; \mbox{diag}(0,1,0) \otimes P_{+y} \; ,
\nonumber \\
B_{-\mu} (x) & = & - B_{+\mu}(x - \hat{\mu})^T\; \; \; \; , 
\; \; \mu = 1,2,3 \; .
\end{eqnarray} 
Using these definitions, the action takes the simple form,
\begin{equation}
S \; = \; \frac{1}{2} \sum_{x,y} \psi(x)^T K(x,y) \; \psi(y) \; ,
\end{equation}
where,
\begin{equation}
K(x,y) \; = \; \widetilde{M} \; \delta(x,y) \; + \; t \sum_{\mu = \pm 1}^{\pm 3}
B_\mu(x) \; \delta(x + \hat{\mu}, y) \; .
\end{equation}
The kernel $K$ is obviously anti-symmetric. With this form for the action, the 
partition function can be partially evaluated, 
\begin{eqnarray}
Z &=& 2^{-3V}\!\!\sum_{ \{ A_{\mu \nu}(x) \} }
\int d \psi \; e^{\frac{1}{2} \psi^T K \psi} =
2^{-3V}\!\!\sum_{\{A_{\mu \nu}(x)\}} \mbox{Pf} K \nonumber\\
&=& 2^{-3V}\!\!\sum_{\{A_{\mu \nu}(x)\}} \sqrt{ \mbox{det} K }
= 2^{-3V}\!\!\sum_{\{A_{\mu \nu}(x)\}} \sqrt{ \mbox{det} \widetilde{M} \; 
\mbox{det}[ 1 + t H] }
\label{3z}
\end{eqnarray}
As in the 2-$D$ case the determinant can be expanded 
(use det$\widetilde{M} = 1$) to give the expansion for the partition function,
\begin{equation}
Z \; = \;  2^{-3V}\!\!\sum_{\{A_{\mu \nu}(x)\}} \exp \left( - \frac{1}{2} 
\sum_{n=1}^\infty \frac{(-t)^n}{n} 
\mbox{Tr}[ H^n] \right) \; ,
\label{3exp}
\end{equation}
where the hopping matrix $H$ takes the form
\begin{equation}
H(x,y) \; = \; \sum_{\mu = \pm 1}^{\pm 3} C_\mu (x) \; 
\delta(x+\hat{\mu}, y) \; .
\end{equation}
The fields $C_\mu(x)$ are obtained from $B_\mu(x)$ by multiplication with 
$\widetilde{M}^{-1}$ 
from the left. Making use of the block-diagonal form of $\widetilde{M}$ 
we find their explicit expression
\begin{eqnarray}
C_{+1} (x) & = & A_{12}(x) \; \mbox{diag}(0,1,0) \otimes H_{+x} +
A_{13}(x) \; \mbox{diag}(0,0,1) \otimes H_{+x},
\nonumber \\
C_{-1} (x)\! & = &\! A_{12}(x-\hat 1) \; \mbox{diag}(0,1,0) \otimes H_{-x} +
A_{13}(x-\hat 1) \; \mbox{diag}(0,0,1) \otimes H_{-x},
\nonumber \\
C_{+2} (x)\! & = &\! A_{12}(x) \; \mbox{diag}(1,0,0) \otimes H_{+x} +
A_{23}(x) \; \mbox{diag}(0,0,1) \otimes H_{+y},
\nonumber \\
C_{-2} (x)\! & = &\! A_{12}(x-\hat 2) \; \mbox{diag}(1,0,0) \otimes H_{-x} +
A_{23}(x-\hat 2) \; \mbox{diag}(0,0,1) \otimes H_{-y},
\nonumber \\
C_{+3} (x)\! & = &\! A_{13}(x) \; \mbox{diag}(1,0,0) \otimes H_{+y} +
A_{23}(x) \; \mbox{diag}(0,1,0) \otimes H_{+y},
\nonumber \\
C_{-3} (x)\! & = &\! A_{13}(x-\hat 3) \; \mbox{diag}(1,0,0) \otimes H_{-y} +
A_{23}(x-\hat 3) \; \mbox{diag}(0,1,0) \otimes H_{-y},
\nonumber \\
\label{cfields}
\end{eqnarray}
The hopping generators $H_{\pm x}, H_{\pm y}$ were already used in the
2-$D$ case (\ref{hmatrices}). To compute the partition function we once 
again must study the traces of powers
of the hopping matrix,
\begin{eqnarray}
\mbox{Tr} [H^n]\!\! & = & \!\!\sum_{x} \sum_{\mu_1, \dots \mu_n}
\delta(x + \hat{\mu}_1 + \dots + \hat{\mu}_n, x) 
\nonumber \\
& & \!\!\mbox{Tr} [ C_{\mu_1}(x) C_{\mu_2}(x + \hat{\mu}_1) 
\dots C_{\mu_n} ( x + \hat{\mu}_1 + \dots + \hat{\mu}_{n-1}) ] .
\label{3htrace}
\end{eqnarray}
The result is similar to the 2-$D$ case but the remaining traces 
now also depend on $x$. Again it can be seen immediately from the Kronecker
delta, that the terms in the sum have support on connected, closed loops 
with base point $x$. The fact that closed loops have even length  
forces Tr$[H^n]$ to vanish for odd $n$.

We remark that since the partition function (\ref{3exp}) is now a sum over 
auxiliary fields a simple computation of the radius 
of convergence of the expansion is not available.  This also prevents a simple
exponentiation of the terms in the statistical sum to obtain the free energy.
For an alternative approach, using different geometric factors see
\cite{orland}.

\subsection{Surfaces}
In this section we will show that the expansion (\ref{3z}), (\ref{3exp}),
(\ref{3htrace}) gives rise to an interpretation in terms of surfaces with
intrinsic geometry
(see also \cite{fradkin}, \cite{itzykson2}, \cite{sedra}-\cite{distler}). 
There is however an intermediate step, which we discuss
first, where the terms in (\ref{3htrace}) are interpreted as loops 
corresponding to ribbons of plaquettes.

It was already mentioned that the loops supporting the contributions
to (\ref{3htrace}) have to be closed and connected. Using essentially the
same arguments as in the 2-$D$ case one can show that the trace over 
the fields $C_\mu(x)$ remains invariant under reversing the orientation of the
loop.
Thus, as in the 2-$D$ case, one can fix an orientation for 
each loop and write a factor 2 in front of the sum in (\ref{3htrace}),
which will later cancel the factor of 1/2 from the square root 
in (\ref{3exp}).

It is crucial to observe, that the special form (\ref{cfields}) 
of the matrix valued 
fields $C_\mu(x)$ reduces the contributions to (\ref{3htrace}) to 
effectively 2-dimensional objects. In particular the matrices 
diag$(1,0,0)$, diag$(0,1,0)$, diag$(0,0,1)$ in (\ref{cfields}) form 
a complete set of projectors, corresponding to the three coordinate planes
of the dual lattice. This has two important implications:

Firstly we note, that the loops in (\ref{3htrace}) are non-back-tracking.
This is a consequence of $H_{\pm \mu} H_ {\mp \mu} = 0$  (compare the
discussion for the 2-$D$ case) and the fact that the mentioned projectors
do not allow for cross-terms in $C_{\pm \mu}(x) C_{\mp \mu} (x \pm \hat{\mu})$ 
(each $C_\mu(x)$ is a sum of two terms).

Secondly, the projectors imply that the choice of just two of the
matrices $C_\mu(x)$ in the trace Tr$[C_{\mu_1}(x) \dots C_{\mu_{2n}} 
(x + \mu_{1} + \dots + \mu_{2n - 1})]$ already fixes a coordinate plane for 
all the terms in this trace. Assume, for example, that the first factor
in the trace is one of $C_{\pm1}(x)$ and any other
factor is one of $C_{\pm 2}(x)$, then from the explicit representation 
(\ref{cfields}) it is clear that only the terms containing the projector
diag$(0,0,1)$ can contribute. One is confined to the 1-2 plane, since 
only the $C_{\pm 1}(x)$ and $C_{\pm 2}(x)$ contain this projector. 
The formula for all terms in this plane (i.e.~all terms associated with the
projector diag$(0,0,1)$) can be read of from (\ref{3htrace}) and 
(\ref{cfields}) and is given by 
\begin{equation}
2 \sum_{x} \sum_{\gamma \in L^{(2n)}_{x,12} }
\mbox{Tr} \Big[ \prod_{\mu \in \gamma} H_\mu \Big] \; 
\prod_{(y,\nu) \in \gamma} A_{\nu 3} (y) \; .
\label{plane12}
\end{equation}
To abbreviate the notation we defined 
$A_{-\nu 3}(y) \equiv A_{\nu 3}( y - \hat{\nu})$ for $\nu = 1,2$. 
The set $L^{(2n)}_{x,12}$ is defined to be
the set of all closed, non-back-tracking 
loops with a chosen orientation and length $2n$ 
in the 1-2 plane, with base point $x$.
In the product under the trace the indices $\mu$ of $H_\mu$ can have 
the values $\pm x, \pm y$ (corresponding to $\pm 1$ and $\pm 2$ directions 
on the dual lattice). 
The trace was already computed in the 2-$D$ case and 
the result is $-(-1)^{n(\gamma)}$. The product over the auxiliary field
picks up all the terms $A_{\nu 3}(x)$ along the loop $\gamma$.

\begin{figure}[htbp]
\epsfysize=1.1in
\hspace*{16mm}
\epsfbox[0 0 443 125] {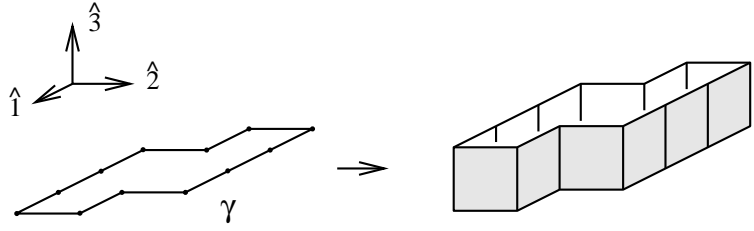}
\caption{{\sl Example for a loop $\gamma$ in the 1-2-plane and the ribbon 
spanned by the links of $\gamma$ together with the unit vector in 
3-direction.}
\label{ribbon}}
\end{figure}

Actually the geometrical objects that appear in the above sum should be thought
of as ribbons of plaquettes. 
These ribbons are defined to be the set of plaquettes
spanned by the links of $\gamma$ together with the unit vector in 
$\hat{3}$-direction. In the above expression (\ref{plane12})
one can interpret the ribbons as being dressed with the  
auxiliary fields $A_{\mu 3}(x)$ defined on the plaquettes of the ribbon.
For the other coordinate planes the ribbons are constructed likewise.
The concept of ribbons is illustrated in Fig.~\ref{ribbon}.
\vspace{3mm}

The contributions to (\ref{3htrace})
in the other two coordinate planes have the same structure as 
(\ref{plane12}), giving
the final result 
\begin{eqnarray}
\mbox{Tr}[H^{2n}] & = &\!\!- 2 \sum_x \Big( 
\sum_{\gamma \in L^{(2n)}_{x,12} }
(-1)^{n(\gamma)}  
\prod_{(y,\nu) \in \gamma} A_{\nu 3} (y) 
\nonumber \\
& + &\!\!\!\! \sum_{\gamma \in L^{(2n)}_{x,13} }
(-1)^{n(\gamma)}  
\prod_{(y,\nu) \in \gamma} A_{\nu 2} (y)  +  
\sum_{\gamma \in L^{(2n)}_{x,23} }
(-1)^{n(\gamma)}  
\prod_{(y,\nu) \in \gamma} A_{\nu 1} (y) \Big) .
\nonumber
\end{eqnarray}
To shorten our notation we identified $A_{\mu \nu} (x) \equiv A_{\nu \mu} (x)$
and introduced $A_{-\mu \nu}(x) \equiv A_{\mu \nu} (x - \hat{\mu})$
(both definitions for $\mu, \nu = 1,2,3; \mu \neq \nu$).

Inserting the result for the trace Tr$[H^{2n}]$
into the expansion (\ref{3exp}) and this 
into the expression (\ref{3z}) for the partition function one ends up with
\begin{eqnarray} 
Z = 2^{-3V} \sum_{\{A_{\mu \nu}(x) = \pm 1\}} \!
& & \! \prod_{x_3} \exp \left(
\sum_{\gamma \in {\cal L}^{con}_{exp}(12,x_3) } 
\frac{(-1)^{n(\gamma)}}{I(\gamma)} \; t^{|\gamma|} \; 
\prod_{(y,\nu) \in \gamma} A_{\nu 3} (y) \right) 
\nonumber \\
& & \! \prod_{x_2} \exp \left(
\sum_{\gamma \in {\cal L}^{con}_{exp}(13,x_2) } 
\frac{(-1)^{n(\gamma)}}{I(\gamma)} \; t^{|\gamma|} \; 
\prod_{(y,\nu) \in \gamma} A_{\nu 2} (y)  \right)
\nonumber \\ 
& & \! \prod_{x_1} \exp \left(
\sum_{\gamma \in {\cal L}^{con}_{exp}(23,x_1) } 
\frac{(-1)^{n(\gamma)}}{I(\gamma)}  \; t^{|\gamma|} \; 
\prod_{(y,\nu) \in \gamma} A_{\nu 1} (y)  \right)\; .
\nonumber \\
\label{3za}
\end{eqnarray}
This formula is a representation of the 3-$D$ Ising model as a product of 
2-$D$ Ising models in their loop representation coupled to an
external field.
We introduced ${\cal L}^{con}_{exp}(12,x_3)$ to be the set of 
loops ${\cal L}^{con}_{exp}$ already discussed in the 2-$D$ case, but 
living in particular in the 1-2 plane of the 3-$D$ lattice 
with 3-coordinate $x_3$.  The sets ${\cal L}^{con}_{exp}(13,x_2)$ and 
${\cal L}^{con}_{exp}(23,x_1)$ are defined likewise. By comparing 
(\ref{3za}) with the 2-$D$ expression (\ref{expsumlocal}) one finds 
that the 3-$D$ partition function is a product of 2-$D$ partition functions
with locally varying Boltzmann factors $t A_{\mu \nu}(x)$. Making use of the
fact that the representations (\ref{expsumlocal}) and (\ref{z2local})
are equal (drop the overall factor in (\ref{z2local})) we obtain
\begin{eqnarray} 
Z = 2^{-3V} \sum_{\{A_{\mu \nu}(x) = \pm 1\}} \!
& & \!\!\!\prod_{x_3} \left(
\sum_{\gamma \in {\cal L}_{int}(12,x_3) } \!\!\! 
(-1)^{n(\gamma)} \; t^{|\gamma|} \; 
\prod_{(y,\nu) \in \gamma} A_{\nu 3} (y) \right) 
\nonumber \\
& & \!\!\!\prod_{x_2} \left(
\sum_{\gamma \in {\cal L}_{int}(13,x_2) } \!\!\! 
(-1)^{n(\gamma)} \; t^{|\gamma|} \; 
\prod_{(y,\nu) \in \gamma} A_{\nu 2} (y)  \right)
\nonumber \\ 
& & \!\!\!\prod_{x_1} \left(
\sum_{\gamma \in {\cal L}_{int}(23,x_1) } \!\!\! 
(-1)^{n(\gamma)} \; t^{|\gamma|} \; 
\prod_{(y,\nu) \in \gamma} A_{\nu 1} (y)  \right)
\nonumber \\ & &
\nonumber \\
=\prod_{x_1,x_2,x_3} 
\sum_{
\stackrel{ \scriptstyle \gamma_{12} \in {\cal L}_{int}(12,x_3) }{
\stackrel{ \scriptstyle \gamma_{13} \in {\cal L}_{int}(13,x_2) }{
\gamma_{23} \in {\cal L}_{int}(23,x_1) }} } \! \! \! & &
t^{|\gamma_{12}|+|\gamma_{13}|+|\gamma_{23}|} \; 
(-1)^{n(\gamma_{12})+n(\gamma_{13})+n(\gamma_{23})}
\nonumber \\
2^{-3V}\!\!\!\!\!\!\sum_{\{A_{\mu \nu}(x) = \pm 1\}}\!\!\!
\!\!\!\prod_{\;\;\;\;(y_{12},\nu_{12}) 
\in \gamma_{12}}\!\!\!\!\!\!A_{\nu_{12} 3}\!\!\!\!\!&\!\!&\!\!\!\!\!\! (y_{12}) 
\!\!\!\!\!\prod_{\;\;\;\;(y_{13},\nu_{13}) 
\in \gamma_{13}}\!\!\!\!\!A_{\nu_{13} 2} (y_{13})
\!\!\!\prod_{\;\;\;\;(y_{23},\nu_{23}) 
\in \gamma_{23}}\!\!\!\!\!A_{\nu_{23} 1} (y_{13}) . 
\nonumber \\
\label{surf1}
\end{eqnarray}
${\cal L}_{int}(12,x_3)$ is the set of loops ${\cal L}_{int}$ defined 
in the 2-$D$ case, but living in the 1-2 coordinate plane with 3-coordinate 
$x_3$. ${\cal L}_{int}(13,x_2)$ and ${\cal L}_{int}(23,x_1)$ are defined
likewise.
The representation (\ref{surf1}) has an interesting interpretation in terms 
of surfaces. The last term in (\ref{surf1}), i.e.~the sum over products 
of $A_{\mu \nu}(x)$, is non-vanishing only if each $A_{\mu \nu}(x)$ 
occurs an even number of times otherwise the sum will cancel terms which
differ only by a minus sign. All non-vanishing contributions to the sum can
be given a geometrical interpretation. This is achieved by associating a ribbon
to every loop (as described earlier), and then dressing the ribbons with the
auxiliary fields $A_{\mu\nu}(x)$ defined on the plaquettes of that ribbon. 
The evenness condition on the auxiliary fields forces non-vanishing 
contributions to come from networks of ribbons that cover each plaquette an
even number of times (or not at all). 
The set of loops in the sum, ${\cal L}_{int}$, was restricted to loops without 
iterated links, hence, the double covering of a plaquette has to come from 
ribbons corresponding to different coordinate planes.
The surfaces are thus built by ribbons
wrapping around the surface, such that each plaquette is covered 
by 2 ribbons corresponding to different coordinate planes. This ensures that
the sum over the $A_{\mu \nu}(x)$ is nonvanishing, in particular it produces 
a factor of $2^{3V}$ which cancels the overall factor in (\ref{surf1}). 

\begin{figure}[htbp]
\epsfysize=0.5in
\hspace*{12mm}
\epsfbox[0 0 444 54] {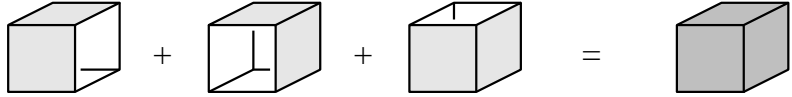}
\caption{ {\sl The three ribbons that build the surface of the cube.
Each face of the cube is covered exactly twice by plaquettes 
from two ribbons corresponding to different coordinate planes.}
\label{cube}}
\end{figure}

\begin{figure}[htbp]
\epsfysize=1.3in
\hspace*{8mm}
\epsfbox[0 0 507 155] {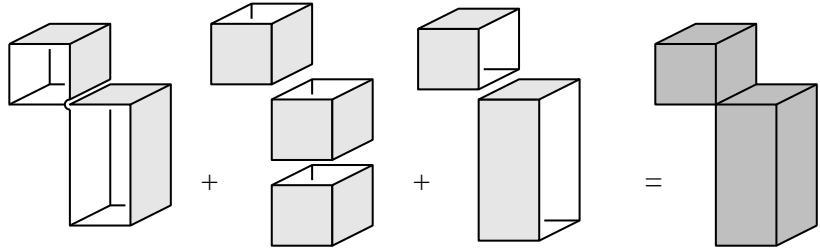}
\caption{ {\sl Example how a surface with one link of self-intersection is
composed from ribbons. Each plaquette of the surface is covered twice 
by ribbons corresponding to different coordinate planes.}
\label{selfintsurf}}
\end{figure}

As an example, we show in Fig.~\ref{cube} how the surface of a simple
cube is generated from three ribbons in the three coordinate planes. In 
Fig.~\ref{selfintsurf} we give an example for a surface which 
self-intersects on one link and show how it is composed from the ribbons.
\vspace{3mm}
\vspace{3mm}

When one sums over all loops in all coordinate planes, all possible
surfaces $\sigma$ on the dual lattice are 
summed over. Since each plaquette is covered twice by the ribbons, the 
factor $t$ is counted twice, giving the overall factor $t^{2|\sigma|}$,
where $|\sigma|$ denotes the area of the surface. The factor  
$(-1)^{n(\gamma_{12})+n(\gamma_{13})+n(\gamma_{23})}$ turns into
$(-1)^{L(\sigma)}$, where $L(\sigma)$ is the number of links where the 
surface self-intersects. The final result for the 3-$D$ partition function is
\begin{equation}
Z \; = \; \sum_{\sigma \in {\cal S}_{int}} 
(-1)^{L(\sigma)} \; t^{2 |\sigma|} \; .
\label{surf2}
\end{equation}
The set ${\cal S}_{int}$ are the surfaces which can be obtained by 
wrapping ribbons corresponding to the subset of loops in 
${\cal L}_{int}$ onto each other. 
More explicitly the surfaces in ${\cal S}_{int}$ are closed, not necessarily 
connected surfaces, which are allowed to self-intersect, 
but plaquettes may not touch each other.

We remark, that when the set of loops in ${\cal L}_{int}$ is extended, so is
the class of allowed surfaces in ${\cal S}_{int}$. For example if 
loops, where links can be occupied several times, are taken into account, 
one can generate surfaces where plaquettes of the surface touch each other. 
However, they are no 
longer necessarily closed surfaces. Consider e.g.~the loop depicted in 
Fig.~\ref{loopsfig}~(c). It gives rise to a ribbon where each of 
the plaquettes is covered twice, and the contribution thus survives 
summing over the auxiliary fields $A_{\mu \nu}(x)$. The corresponding 
surface has however a boundary. Thus as in the 2-$D$ case it is possible
to allow for a larger class of surfaces than $S_{int}$. 
Many of these surfaces, such as our example
of the doubly covered ribbon have no counterparts in the set ${\cal S}_{ext}$  
of surfaces with extrinsic geometry and get cancelled. 
All these more general sets of surfaces can be generated by allowing for 
generalized sets of loops ${\cal L}_{int}$. In other words, the surfaces 
are classified by the behavior of the loops which are obtained as an 
intersection of the surfaces with all possible coordinate planes on the 
lattice. This makes (\ref{surf1}) and 
(\ref{surf2}) very powerful for a simple characterization of possible 
classes of surfaces with intrinsic geometry. For all 
these cases the geometry factor is $(-1)^{L(\sigma)}$ and the partition 
function is given by (\ref{surf2}).

\subsection{Correlation functions in terms of surfaces}
In this section we derive a simple formula for 2$n$-point functions in terms
of surfaces. We will compute these correlation functions by making use of 
the variable bond Ising model,
\begin{equation}
Z\Big(\{ \beta_{\langle x,y \rangle}\}\Big) \; = \; 
\sum_{\{ s(x) = \pm 1\}} \exp \left( 
\sum_{\langle x, y \rangle} \beta_{\langle x,y \rangle} s(x) s(y) \right) \; .
\end{equation}
This expression can serve as a generating functional for 2-point functions
(or more general 2n-point) functions: Let 0 and z be the two lattice sites
where the spins in the 2-point function are located. Connect them by an 
arbitrary set of links, $\Gamma$, forming a path from $0$ to $z$. It is 
convenient to encode the path $\Gamma$ by the set of directions $\hat{\mu}_1 
\dots \hat{\mu}_{|\Gamma|}$ which trace it out when starting at 0 
(here $|\Gamma|$ denotes the length of $\Gamma$.) Then one can write, 
\begin{eqnarray}
\langle s(0) s(z) \rangle & = & 
\langle s(0) 
\underbrace{s(\hat{\mu}_1) s(\hat{\mu}_1)}_{=1} 
\dots
\underbrace{s(z - \hat{\mu}_{|\Gamma|}) s(z - \hat{\mu}_{|\Gamma|})}_{=1} 
s(z) \rangle =
\nonumber \\
\Big\langle \prod_{\langle x, y \rangle \in 
\Gamma}\!\!s(x)s(y) \Big\rangle & = &
\frac{1}{Z} \left( \prod_{\langle x, y \rangle \in \Gamma}
\frac{\partial}{\partial \beta_{\langle x, y \rangle}}\right)
Z\Big(\{ \beta_{\langle x, y \rangle}\}\Big) 
\Big|_{\beta_{\langle x, y \rangle} = \beta} \; .
\nonumber \\
\label{generating}
\end{eqnarray}
We remark that it is not necessary to set the coupling constants equal 
(this was done only for notational convenience), and all results below can
easily be generalized to the variable bond model. Also, the 
generalization to 2$n$-point functions is straightforward: simply replace the 
single path $\Gamma$ by a network of paths, consisting of links on the lattice,
which connect all sites occuring in the 2$n$-point function. We will show in 
the end, that the result is independent of the choice of the network.

The essential step is to replace the partition function 
of the variable 
bond model by the corresponding expression in terms of surfaces. 
For notational convenience we will work with the representation in terms of 
surfaces with extrinsic geometry rather than the more cumbersome intrinsic 
geometry. However, it is of course possible to express the final result in
terms of surfaces with intrinsic geometry.
For the variable bond model the surface representation is given by,
\begin{equation}
Z\Big(\{\beta_{\langle x, y \rangle}\}\Big) \; = \; 
2 \exp \left( \sum_{\langle x, y \rangle} \beta_{\langle x, y \rangle} \right)
\sum_{\sigma \in {\cal S}_{int}} \;
\prod_{\langle x, y \rangle: \; \langle x, y\rangle \in \sigma^*} 
e^{-2 \beta_{\langle x, y \rangle} }  \; .
\label{randsurf}
\end{equation}
Here $\sigma^*$ denotes the set of links dual to the plaquettes of $\sigma$.
A straightforward but lengthy calculation gives
\begin{eqnarray}
& & \prod_{\langle x, y \rangle \in \Gamma}
\frac{\partial}{\partial \beta_{\langle x, y \rangle}}
Z\Big(\{ \beta_{\langle x, y \rangle}\}\Big) 
\Big|_{\beta_{\langle x, y \rangle}} \; = \; Z \; + 
\nonumber \\
& & 2 e^{\beta 3 V} (-2) \; \sum_{\langle x, y \rangle \in \Gamma} \; \; \;
\sum_{\{\sigma: \; \langle x, y \rangle \in \sigma^* \} }
\!\!\!\!\!\!\! e^{-2 \beta |\sigma|} 
\qquad +
\nonumber \\
& & 2 e^{\beta 3 V} (-2)^2  
\sum_{\langle x,y \rangle < \langle x^\prime,y^\prime \rangle \in \Gamma} \; \;
\sum_{\{\sigma: \; \langle x, y \rangle,
\langle x^\prime,y^\prime \rangle \in \sigma^* \}} 
\!\!\!\!\!\!\!\!\!\!\!\!\! e^{-2 \beta |\sigma|} \qquad +
\nonumber \\
& & 2 e^{\beta 3 V} (-2)^3  \!\! 
\sum_{\langle x, y \rangle < \langle x^\prime,y^\prime\rangle < 
\langle x^{\prime \prime},y^{\prime \prime} \rangle \in \Gamma} \; \; \; \;
\sum_{\{\sigma: \; \langle x, y \rangle, \langle x^\prime,y^\prime \rangle,
\langle x^{\prime \prime},y^{\prime \prime} \rangle \in \sigma^*\}} 
\!\!\!\!\!\!\!\!\!\!\!\!\!\!\!\!\!\!\!\!\!\!\!
e^{-2 \beta |\sigma|} \quad +
\nonumber \\
& & \dots \dots
\nonumber \\
& & 2 e^{\beta 3 V} (-2)^{|\Gamma|} 
\sum_{\{\sigma: \; |\Gamma \cap \sigma^*| = |\Gamma| \}} 
\!\!\! e^{-2 \beta |\sigma|} \; .
\label{longform}
\end{eqnarray}
We introduced an (arbitrary) ordering $\langle x, y \rangle
 < \langle x^\prime, y^\prime \rangle \dots$
of the links in $\Gamma$. In the last term we introduced 
$|\Gamma \cap \sigma^*|$ to be the number of links in $\Gamma$ which are 
also in $\sigma^*$. The final step is to interpret all the 
contributions in terms of surfaces. Let $\sigma$ be a surface that 
intersects $\Gamma$ at $m$ links, then $\sigma$ 
is counted in the first $m+1$ terms on 
the right hand side of (\ref{longform}). To begin with, 
$\sigma$ is certainly counted in 
$Z$, which is the first term on the right hand side of (\ref{longform}),
with a factor $1=(-2)^0 {m \choose 0}$. Of course, there is also the factor 
$2 \exp(\beta[
3V - 2|\sigma|])$ but this will appear in all terms in (\ref{longform}), 
so we will neglect it for the moment. 
In the second term in (\ref{longform}) the surface $\sigma$ gets firstly an 
overall factor of $-2$ and then it is counted $m$-times, since the first 
sum in this term runs over all links. Thus, in the second term of 
(\ref{longform}) $\sigma$ acquires the factor $(-2) m =
(-2) {m \choose 1}$. 
In the third term the overall factor is $(-2)^2$ and since the links 
in the sum are ordered the surface occurs ${m \choose 2}$ times. 
Thus the factor is $(-2)^2 
{m \choose 2 }$. In general a surface which intersects $\Gamma$ 
at $m$ links obtains a factor of $(-2)^{i-1} {m \choose i-1}$ from the $i$-th
term in (\ref{longform}). These factors can be summed up to give
\[
\sum_{i=1}^{m+1} {m \choose i-1} (-2)^{i-1} \; = \; 
(1-2)^m \; = \; (-1)^m \; .
\]
We thus end up with the surprisingly simple result 
(the overall factor $2 e^{\beta 3 V}$ gets cancelled when dividing by $Z$)
\begin{equation}
\langle s(x_1) s(x_2) \dots s(x_{2n}) \rangle \; = \; 
\sum_{\sigma \in {\cal S}_{ext}} 
(-1)^{|\Gamma \cap \sigma^*|} \; \; t^{2 |\sigma|}
\; \Bigg/ \; \sum_{\sigma \in {\cal S}_{ext}} 
t^{2 |\sigma|} \; ,
\label{npoint}
\end{equation}
where $\Gamma$ is an arbitrary network of paths connecting the sites
$x_1, \dots x_{2n}$ and $t$ denotes $\exp(-\beta)$.
It is simple to see the independence of (\ref{npoint}) from the 
choice of $\Gamma$: The network $\Gamma$ can be deformed arbitrarily
by adding single plaquettes to $\Gamma$ and since a plaquette always 
intersects an even number of times (0, 2 or 4-times) 
with a given surface, the intersection term
$(-1)^{|\Gamma \cap \sigma^*|}$ remains invariant under such deformations. 
We finally remark, that the same formula holds if one works with the 
representation in terms of surfaces with intrinsic geometry (certainly
the self-intersection factor $(-1)^{L(\sigma)}$ has to be re-inserted).
The formula also holds in 2-$D$, 
when the surfaces $\sigma$ are replaced by the 
loops $\gamma$.

\section{Discussion}
In this paper we discussed the hopping expansion of the Grassmann
representation of the Ising model in both 2 and 3 dimensions. 
For the 3-$D$ case an auxiliary field was introduced in order 
to write the action in quadratic form. 
In both dimensions the expansion was interpreted in terms of
loops or surfaces. It was shown that the hopping expansion, with
its calculus of hopping generators, provides an exact 
algebraic representation of the
expansions in terms of loops and surfaces, respectively. This connection allows
for a simple algebraic treatment of many of the problems
which emerge in the loop or surface expansion and are rather intractable 
when working with the original geometrical objects.

Counting problems and symmetry factors can be analyzed in
a rather simple fashion using the algebraic representation. As an application, 
in 2-$D$, we gave the corrected result of the loop representation of the free 
energy and computed its radius of convergence showing that it is 
determined by the critical temperature. In 3-$D$ we derived a representation 
of the partition function as a product of 2-$D$ Ising models 
in their loop
representation coupled to an auxiliary field.  We gave a simple proof that the 
self intersection factor leads to the cancellations necessary so that 
the sum over surfaces with intrinsic geometries reproduces the correct 
partition function. The possible classes of surfaces to be summed over 
were characterized by their intersections with all coordinate planes of the
lattice. Our formula for the $2n$-point functions in terms of surfaces
illustrates the tight relation between dynamics and geometry in 2- and 
3-dimensional Ising models.
\\ 

The fermionic representation of the Ising model
is also known in dimensions higher than 3.
It involves however terms of order larger than the quartic contributions
generating the plaquettes in 3-$D$. By introducing several auxiliary
fields one could write the action as a quadratic form and perform 
the hopping expansion along the lines of this paper. However, the geometrical 
interpretation of the emerging structures might turn out to be rather 
involved.

The results obtained here might have a more or less straightforward 
generalization to other models where Grassmann representations exist
\cite{samuel1}, such as vertex models and the Ashkin-Teller model. It would be 
interesting to work out representations of these models in terms of
geometrical objects with intrinsic geometry. In particular 
representations of $n$-point functions in terms of these surfaces
could be analyzed further. The outcome of such an enterprise might be 
a better understanding of the relation between dynamics and geometrical
concepts.
\\
\\
{\bf Acknowledgement:}
This work is supported in part by NSERC of
Canada, NATO CRG 970561 and the Fonds zur F\"orderung der
Wissenschaftlichen Forschung in \"Osterreich under project number
J1577-PHY.

\end{document}